\documentclass[12pt,draftclsnofoot,onecolumn]{IEEEtran}

\usepackage{amssymb}
\usepackage[cmex10]{amsmath}
\usepackage{stfloats}
\usepackage{graphicx}
\usepackage{subfigure}
\usepackage{tabularx}
\usepackage{epsfig,epsf,color,balance,cite}
\usepackage{verbatim}
\usepackage{url}
\usepackage{bm}

\newtheorem{theorem}{\textbf{Theorem}}
\newtheorem{lemma}{\textbf{Lemma}}

\usepackage{algorithm}
\usepackage{algpseudocode}
\begin{document}

\title{Cache Placement in Two-Tier HetNets with Limited Storage Capacity: Cache or Buffer?}
\author{
\IEEEauthorblockN{Zhaohui Yang, 
Cunhua Pan, \IEEEmembership{Member, IEEE},
Yijin Pan,
Yongpeng Wu, \IEEEmembership{Senior Member, IEEE},
Wei Xu, \IEEEmembership{Senior Member, IEEE},
 and
Ming Chen, \IEEEmembership{Member, IEEE}
}
\thanks{Z. Yang, Y. Pan, W. Xu, and M. Chen are with the National Mobile Communications Research Laboratory, Southeast University, Nanjing 211111, China, Email: \{yangzhaohui, panyijin, wxu, chenming\}@seu.edu.cn.}
\thanks{C. Pan is with School of Electronic Engineering and Computer Science, Queen Mary University of London, London E1 4NS, U.K,
 (Email: c.pan@qmul.ac.uk).
}
\thanks{Y. Wu is with the Department of Electronic Engineering, Shanghai
Jiao Tong University, China, Minhang 200240, China (Email:
yongpeng.wu@sjtu.edu.cn).
}
}

\maketitle
\begin{abstract}
In this paper, we aim to minimize the average file transmission delay via bandwidth allocation and cache placement in two-tier heterogeneous networks with limited storage capacity, which consists of cache capacity and buffer capacity. For average delay minimization problem with fixed bandwidth allocation, although this problem is  nonconvex, the optimal solution is obtained in closed form according to the Karush-Kuhn-Tucker conditions. To jointly optimize bandwidth allocation and cache placement, the optimal bandwidth allocation is first derived and then substituted into the original problem. The structure of the optimal caching strategy is presented, which shows that it is optimal to cache the files with high popularity instead of the files with big size. Based on this optimal structure, we propose an iterative algorithm with low complexity to obtain a suboptimal solution, where the closed-from expression is obtained in each step. Numerical results show the superiority of our solution compared to the conventional cache strategy without considering cache and buffer tradeoff in terms of delay.

\end{abstract}


\IEEEpeerreviewmaketitle
\newpage
\section{Introduction}

To accommodate the growing demand for high data rate transmission and seamless coverage in wireless communications, heterogeneous deployment has been proposed as an effective network architecture \cite{Ghosh2012HetNet}.
In heterogeneous networks (HetNets), small base stations (BSs) are deployed to offload the traffic in high user density area.
To further improve the transmission rate and decrease latency for users, wireless caching is a promising solution by caching popular contents at the network edge \cite{4509746,Wang2014Cache,Sheng2016Enhance,Lu2016Caching,Ali2014Fundamental,7994891,7537172,8761881}.
The recent contributions about cache-aided wireless networks can be classified into two categories: analyzing the content delivery performance and designing cache placement strategies.

Content delivery performance analysis is crucial in reveling the benefits of distributed cache placement in cache-enabled networks \cite{Ji2015Out,Jeon2015Caching,Afzal2016on,Tao2016Content,Stochastic2016,Zhou2017Optimal}.
The throughput-outage tradeoff was investigated in \cite{Ji2015Out} for one-hop device-to-device (D2D) networks, which showed that the user throughput is proportional to the fraction of cached information.
For multi-hop D2D networks, the multi-hop capacity scaling laws were investigated in \cite{Jeon2015Caching}.
It was further shown in \cite{Jeon2015Caching} that a multi-hop transmission
provides a significant throughput gain over one-hop direct transmission for Ad-Hoc networks with cached users.
For cache-enabled cellular networks with coordinated D2D communications, cellular and D2D coverage probabilities were derived in \cite{Afzal2016on}.
In \cite{Tao2016Content,Stochastic2016,Zhou2017Optimal}, multicast beamforming was investigated for cache-enabled content-centric networks.

Cache placement strategies should be properly designed due to the features of link connectivity and channel quality \cite{Ahlehagh2014Video,Wang2016Mob,7805409}.
There are mainly two issues addressed in cache placement problems: the hit probability maximization \cite{Wen2017ICC,Wen2017Cache,Pan2017On,8125744} and the average delay minimization \cite{Shan2013Femto,Cui2016CachingGC,liu2017optimizing,zhang2017cooperative,Xiang2017Cross}.
The hit probability is defined as the probability that a user will find the file he/she is asking for in the cache of the BS he/she is covered from \cite{Blaszczyszyn2015Cache}.
Considering both small BS caching and cooperation in a downlink HetNet,
\cite{Wen2017ICC} first derived a tractable expression for the hit probability by using stochastic geometry, and then optimized the caching distribution to maximize this hit probability.
In \cite{Wen2017Cache}, the structure of the optimal content-placement policies to maximize the hit probability for HetNets was investigated.
The impact of file preference and user willingness on hit probability was investigated in \cite{Pan2017On} by optimizing the cache placement strategy.
Instead of maximizing hit probability, \cite{Shan2013Femto} analyzed the optimal way of assigning files to the small BSs to minimize the average transmission delay.
Joint caching and user association was considered in \cite{Cui2016CachingGC} to minimize the delay for satisfying the transmission demands in cached-enabled HetNets with wireless backhaul.
By exploiting user preference and spatial locality, \cite{liu2017optimizing} investigated the optimal cache policy to minimize the average file download time in HetNets.
In user-centric networks,
the delay-optimal cooperative edge caching was investigated in \cite{zhang2017cooperative}.
Due to limited storage, some files are always not cached in the edge nodes and the delivery of uncached files constitutes a performance bottleneck caused by the buffer, which serves as a short-term memory to temporally store the data.
However, the above contributions \cite{Wen2017ICC,Wen2017Cache,Pan2017On,8125744,Shan2013Femto,Cui2016CachingGC,liu2017optimizing,zhang2017cooperative} all ignored the effect of buffer.
Considering the maximal buffer capacity constraint,
\cite{Xiang2017Cross} investigated the average delay minimization in cache- and buffer-enabled relaying networks.
The maximal buffer capacity is assumed to be fixed in \cite{Xiang2017Cross}, even though considering cache capacity and buffer capacity tradeoff can further improve the system performance.

In this paper, we aim to minimize the average file transmission delay through bandwidth allocation and cache placement in two-tier HetNets with limited storage capacity.
Different from \cite{Xiang2017Cross}, we consider the sum cache and buffer capacity constraint to balance cache and buffer.
The contributions of this paper are summarized as follows:
\begin{enumerate}
  \item  The average file transmission delay minimization problem is formulated by considering transmission delay, fronthaul delay and buffer delay, which reflect the tradeoff of cache and buffer.
      Specifically, the average file delay expression is derived by modeling the distribution of users as independent poisson point process (PPP).
  \item  For cache placement with fixed bandwidth allocation, we have successfully derived the structure of the optimal solution to this nonconvex problem by solving Karush-Kuhn-Tucker (KKT) conditions.
      Based on the optimal structure, the optimal cache strategy can be obtained by comparing finite potentially optimal solutions.
      In this case, the optimal cache strategy indicates the optimal tradeoff of cache capacity and buffer capacity.
  \item For joint bandwidth allocation and cache placement, the optimal bandwidth allocation is first derived in closed form.
      Then, the original problem is transformed into an equivalent problem with respect to the cache strategy by substituting the optimal bandwidth allocation.
      To solve the equivalent nonconvex problem, an iterative algorithm with low complexity is proposed to obtain a suboptimal solution.
\end{enumerate}

The rest of the paper is organized as follows.
In Section $\text{\uppercase\expandafter{\romannumeral2}}$, we introduce the system model.
Problem formulation and analysis are presented in Section $\text{\uppercase\expandafter{\romannumeral3}}$.
Optimal cache placement with fixed bandwidth allocation and joint bandwidth and cache optimization are addressed in Section $\text{\uppercase\expandafter{\romannumeral 4}}$ and Section $\text{\uppercase\expandafter{\romannumeral 5}}$, respectively.
Some numerical results are shown in Section $\text{\uppercase\expandafter{\romannumeral6}}$
and conclusions are finally drawn in Section $\text{\uppercase\expandafter{\romannumeral7}}$.

%


\section{System Model}
\subsection{System Model}
\begin{figure}
\centering
\includegraphics[width=5.2in]{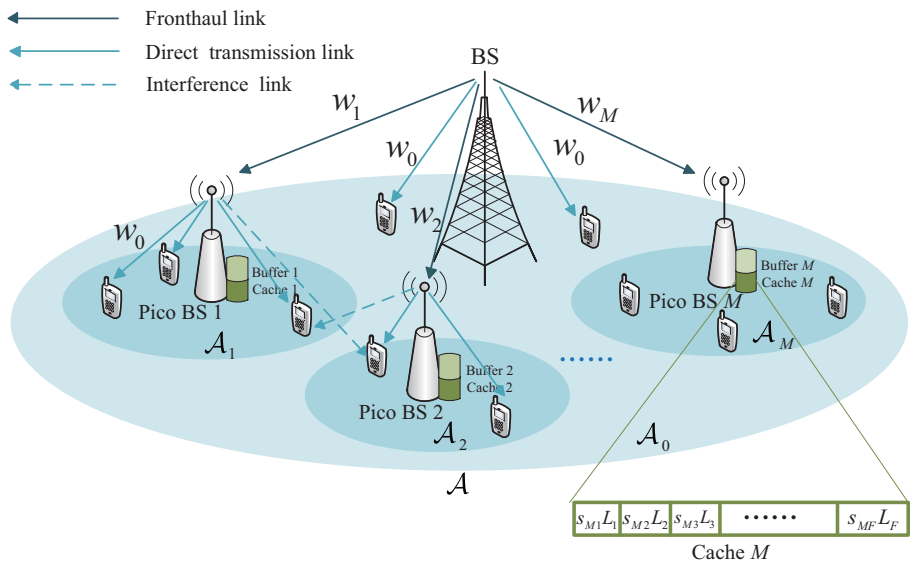}
\vspace{-2em}
\caption{System model.}\label{sys1}
\end{figure}

Consider a cache-enabled HetNet with one macro BS and $M$ pico BSs as shown in Fig. \ref{sys1}.
The set of pico BSs is denoted  by $\mathcal M\triangleq\{1, 2, \cdots, M\}$.
Denote the total coverage area of the macro BS by $\mathcal A$, while the coverage area of pico $m$ is denoted by $\mathcal A_m$.
The coverage area $\mathcal A_m$ ($\mathcal A$) is a circle area centered at pico BS $m$ (macro BS) with radius $r_m$ ($r_0$).
The pico BSs' coverage areas are disjoint, i.e., $\mathcal A_m\cap\mathcal A_n=\varnothing$ for $m, n\in \mathcal M$ and $m\neq n$.
Let ${\mathcal A}_{0}\triangleq\mathcal A\setminus\cup_{m\in\mathcal M}\mathcal A_m$ denote  the set of areas only covered by the macro BS.

The macro BS and all the pico BSs share the same bandwidth $w_0$ for wireless information transmission to users.
The wireless fronthaul links from the macro BS to the pico BSs are assumed to be orthogonal, and denote $w_m$ as the bandwidth allocated to the wireless fronthaul link from the macro BS to pico BS $m$.
As a result, we have
\begin{equation}\label{sys1eq0}
\sum_{m\in\mathcal M\cup\{0\}}w_m \leq W,
\end{equation}
where $W$ is the total bandwidth of the network.

Let $\mathcal F\triangleq\{1,\cdots, F\}$ denote the set of $F$ files.
The length of file $f$ is denoted by $L_f>0$ (measured in bits).
The file popularity distribution $\{q_1,  \cdots, q_F\}$ is assumed to be identical for all users, where $q_f\in(0,1]$ is the popularity of file $f$ and $\sum_{f\in\mathcal F}q_f=1$.
Without loss of generality, the files are sorted as $q_1 > q_2 >\cdots > q_F>0$.

All files are stored at the macro BS, while each pico BS can only cache a subset of the total files due to limited storage capacity.
Denote the storage capacity of pico BS $m$ by $C_m$ (measured in bits).
Assume that each file is further encoded via rateless maximum distance separable codes \cite{Shanmugam2013FemtoCaching}.
Letting $s_{mf}\in[0,1]$ denote the fraction of file $f$ cached at pico BS $m$, we have
\begin{equation}\label{sys1eq1}
\sum_{f\in\mathcal F} s_{mf} L_f \leq C_m.
\end{equation}
It is assumed that $C_m<\sum_{f\in\mathcal F}L_f$, i.e., each pico BS cannot cache all files.
Note that the storage capacity contains both cache capacity and buffer capacity, since cache chip and buffer chip are interchangeable \cite{7927052,kang2012flash,huang2016improving,jiang2002lirs}.
Cache is a long-term memory to cache popular files in a long time, while buffer is a short-term memory to temporally store the file.
Consequently, the remaining part $C_m-\sum_{f\in\mathcal F}s_{mf} L_f$ means the buffer capacity of pico BS $m$.

\begin{figure}
\centering
\includegraphics[width=3.7in]{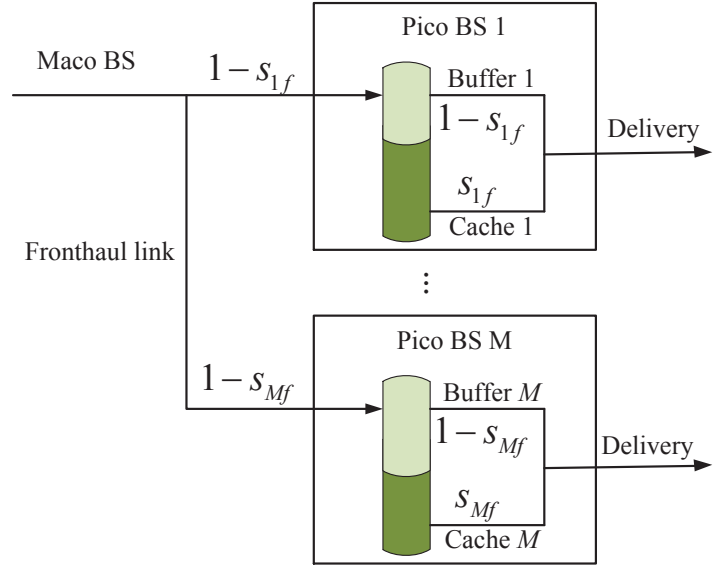}
\caption{The structure of cache and buffer in each pico BS.}\label{sys2}
\end{figure}

The distribution of users in the whole area $\mathcal A$ is modeled as independent PPPs with density $\lambda$.
The users located in ${\mathcal A}_0$ are only served by the macro BS.
As shown in Fig.~\ref{sys2},
for each file $f$, user $i$ located in $\mathcal A_m$ covered by pico BS $m$ first fetches $s_{mf}$ fraction of cached file $f$ from pico BS $m$.
The remaining $1-s_{mf}$ fraction of file $f$ is delivered to pico BS $m$ from the macro BS via the wireless fronthaul link and then relayed to user $i$ from pico BS $m$ with the aid of buffer.

\subsection{Delay Model}
For a user $i$ with location $\xi\in\mathcal A_m$ ($\mathcal A_0$), the file transmission rate from pico BS $m$ (the macro BS) to user $i$ is given by
\begin{equation}\label{sys2eq1}
R_{mi}(\xi)=w_{mi}\log_2\left(1+
\frac{P_m |h_{mi}|^2 (d_{mi}(\xi))^{-\alpha}}
{\sigma^2+\sum_{n\in\mathcal M\cup\{0\}\setminus \{m\} }P_n |h_{ni}|^2 (d_{ni}(\xi))^{-\alpha}}\right), \quad \forall m \in \mathcal M\cup\{0\},
\end{equation}
where $w_{mi}$ is the allocated bandwidth for user $i$ by pico BS $m\in\mathcal M$ (the macro BS for $m=0$),
$\sigma^2$ is the noise power,
$P_m$ is the transmission power of pico BS  $m$ for $m \in\mathcal M$ (the macro BS for $m=0$),
$|h_{mi}|^2\sim\exp(1)$, which is a exponentially distributed random variable with unit mean, denotes the small-scale fading channel gain between user $i$ and pico BS $m$ for $m\in\mathcal M$ (the macro BS for $m=0$),
$d_{mi}(\xi)$ is the distance between user $i$ located in $\xi$ and pico BS $m$ for $m\in\mathcal M$ (the macro BS for $m=0$),
and $\alpha$ is the pathloss exponent.
Since users follow the same PPP distribution, equal bandwidth allocation is adopted in each BS, i.e.,
\begin{equation}\label{sys2eq2}
w_{mi}=\frac{w_0} {U_m}, \quad \forall m \in \mathcal M\cup\{0\},
\end{equation}
where $U_m$ is the number of users located in $\mathcal A_m$.

For fronthaul link, the file transmission rate from the macro BS to pico BS $m$ is
\begin{equation}\label{sys2eq3}
R_{m}=w_{m}\log_2\left(1+
\frac{P_0 |h_{m}|^2 d_{m}^{-\alpha}}
{\sigma^2}\right), \quad \forall m \in \mathcal M,
\end{equation}
where $|h_{m}|^2\sim\exp(1)$ denotes the small-scale fading channel gain between the macro BS and pico BS $m$,
and
$d_{m}$ is the distance between the macro BS and pico BS $m$.

When the file size $L_f$ is large for each file $f\in\mathcal F$ and the duration time of each slot $\delta$ is so small that the needed number of time slots $N$ to download a file is large \cite{liu2017optimizing}, the download time per unit bit for a user located in $\mathcal A_m$ is
\begin{equation}\label{sys2eq3_1}
\lim_{N\rightarrow\infty}
\frac{N\delta }{\sum_{n=1}^NR_{mi}(\xi,n)\delta}
=
\frac1{\lim_{N\rightarrow\infty}\frac1N\sum_{n=1}^NR_{mi}(\xi,n) }
=\frac1{\mathbb E_{U_m,h_{0i},\cdots,h_{Mi},\xi}R_{mi}(\xi) },
\end{equation}
where $R_{mi}(\xi,n)$ is the file transmission rate in the $n$-th time slot.
As a result, the average delay for a user located in $\mathcal A_m$ covered by pico BS $m$ to download file $f$ can be modeled as
\begin{equation}\label{sys2eq3_2}
d_{mf}= \underbrace{{\frac {L_f}{\mathbb E_{U_m,h_{0i},\cdots,h_{Mi},\xi}R_{mi}(\xi)}}}_{\text{transmission delay from pico BS $m$}}
+\underbrace{{\frac {(1-s_{mf})L_f}{\mathbb E_{h_{m}}R_{m}}}}_{\text{fronthaul delay from the macro BS}}
+\underbrace{\frac{(1-s_{mf})L_f}{C_m-\sum_{f\in\mathcal F}s_{mf} L_f}D}_{\text{buffer delay from pico BS $m$}},
\end{equation}
where $D$ is the buffer delay per time.
The third term in (\ref{sys2eq3_2}) denotes the buffer time consumed at pico BS $m$.
With capacity $\sum_{f\in\mathcal F}s_{mf}L_f$ allocating to cache files, the remaining buffer capacity $C_m-\sum_{f\in\mathcal F}s_{mf}L_f$ is only used to relay file from the macro BS to the served user.
To deliver file $f$ with capacity $(1-s_{mf})L_f$, the average portion of time required for delivering file $f$ in the buffer is  $\frac{(1-s_{mf})L_f}{C_m-\sum_{f\in\mathcal F}s_{mf} L_f}$.
According to Little's law \cite{little1961proof} and \cite{Islam2013TWC}, the average delay for file $f$ in the buffer of pico BS $m$, i.e., the average time that a packet is stored in the buffer, can be given by $ \frac{(1-s_{mf})L_f}
{C_m-\sum_{f\in\mathcal F}s_{mf}L_f} D$.

\begin{figure}
\centering
\includegraphics[width=3.6in]{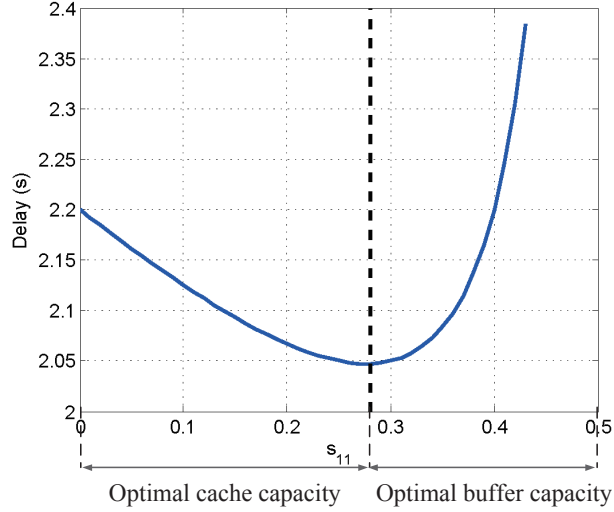}
\caption{Delay $d_{11}$ versus cache placement $s_{11}$ for pico BS 1 with $L_1=1$ Mbits, $C_1=0.5$ Mbits, ${\mathbb E_{U_1,h_{0i},\cdots,h_{Mi},d_{1i}(\xi)}R_{1i}(\xi)}=1$ Mbps, ${\mathbb E_{h_{1}}R_{1}}=1$ Mbps, $D=0.1$ s.}\label{sys3}
\end{figure}

From (\ref{sys2eq3_2}), it is observed that the fronthaul transmission delay decreases with $s_{mf}$, while the buffer delay increases with $s_{mf}$ for $L_f<C_m-\sum_{l\in\mathcal F\setminus\{f\}}s_{ml}L_l$ and decreases with $s_{mf}$ for $L_f\geq C_m-\sum_{l\in\mathcal F\setminus\{f\}}s_{ml}L_l$.
Fig.~\ref{sys3} shows the delay given in (\ref{sys2eq3_2}) for the special case of one cached file in pico BS 1 versus the fraction of cached file.
It can be seen that the average delay first decreases and then increases with the fraction $s_{11}$ of cached file.
This can be explained as follows.
For small $s_{11}$ (less than 0.2),  more bits will be transferred from the macro BS to pico BS 1, which incurs larger fronthaul delay.
When $s_{11}$ becomes large (large than 0.3), the buffer capacity in pico BS is small, which causes large time cost for buffering.

For file $f$, the average delay for a user located in $\mathcal A_0$ served by the macro BS is
\begin{equation}\label{sys2eq5}
d_{0f}= \frac {L_f}{\mathbb E_{U_0,h_{0i},\cdots,h_{Mi},\xi}R_{0i}(\xi)}.
\end{equation}
As a result, the average file transmission delay is given by
\begin{equation}\label{sys2eq6}
D_{\text{avg}}=\sum_{m\in\mathcal M\cup0}\sum_{f\in\mathcal F}q_f d_{mf}.
\end{equation}

\section{Problem Formulation and Analysis}

Based on the above system model, we formulate the joint bandwidth allocation and cache placement problem to minimize the average file transmission delay.
Then, we provide the expressions for average transmission rate.

\subsection{Problem Formulation}
Based on (\ref{sys1eq1})-(\ref{sys2eq6}), the average file transmission delay optimization problem is formulated as
\begin{subequations}\label{PFmin1}
\begin{align}
\mathop{\min}_{ \pmb{s},\pmb w}& \:\;
\sum_{m\in\mathcal M} \sum_{f\in\mathcal F}q_f\left(\frac{ a_m L_f}{w_0}+
\frac{b_m(1-s_{mf})L_f}{w_m} +\frac{(1-s_{mf})L_fD}
{C_m-\sum_{f\in\mathcal F}s_{mf}L_f}
\right)
+\sum_{f\in\mathcal F}q_f\frac{a_0L_f}{w_0}
\\
\textrm{s.t.}
& \:\;\sum_{f\in\mathcal F}s_{mf}L_f\leq C_m, \quad \forall m \in \mathcal M\\
& \:\; \sum_{m\in\mathcal M\cup\{0\}} w_m \leq W\\
& \:\; 0\leq s_{mf}\leq 1,\quad\forall m\in\mathcal M, f \in \mathcal F\\
& \:\; w_m\geq0,\quad\forall m\in\mathcal M\cup\{0\},
\end{align}
\end{subequations}
 where $\pmb s=[s_{11},\cdots,s_{1F},\cdots,s_{MF}]^T$,
 $\pmb w=[w_0,w_1,\cdots,w_M]^T$,
$a_{m}={\frac {w_0}{\mathbb E_{U_m,h_{0i},\cdots,h_{Mi},\xi }R_{mi}(\xi)}}$, $b_m=\frac{w_m}{\mathbb E_{h_m}R_{m}}$, $a_{0}={\frac {w_0}{\mathbb E_{U_0,h_{0i},\cdots, h_{Mi}, \xi}R_{0i}(\xi)}}$, $\forall m \in \mathcal M$.
Constraints (\ref{PFmin1}b) reflect the limitation of the storage capacity,
and constraint (\ref{PFmin1}c) shows that the bandwidth of the system is constrained.
The optimization of cache placement $\pmb s$ is to balance the cache capacity and buffer capacity.
Increasing cached files can improve the file hit ratio, and the average delay for cached files can be reduced, while the average delay for uncached files is increased due to small buffer capacity.
The optimization of bandwidth allocation takes the tradeoff between the resource and traffic demand into consideration, due to the fact that different cache placements lead to different traffic demands among BSs.

There are two difficulties to solve Problem (\ref{PFmin1}).
The first one is to obtain the expressions of parameters $a_m$ and $b_m$, which are determined by the randomness of the number and locations of users as well as channel gains.
The second one is that cache placement variable $\pmb s$ and bandwidth allocation variable $\pmb w$ are coupled in the objective function (\ref{PFmin1}a), which makes Problem (\ref{PFmin1}) a nonconvex problem.

To deal with the first difficulty, we analyze the average delay based on the exponential distribution of channel gains in the following subsection.
As for the second difficulty, we obtain the optimal cache placement in closed form with fixed bandwidth allocation, and provide a low-complexity algorithm to solve the joint bandwidth allocation and cache placement problem.

\subsection{Average Delay Analysis}

\begin{lemma}
The average download time multiplied by bandwidth per bit of the user when
downloading file from pico BS $m\in\mathcal M$ (or the macro BS with $m=0$)
can be obtained as
\begin{eqnarray}\label{PFeq1}
a_m&&\!\!\!\!\!\!\!\!\!\!= \left[\kappa_m \int_0^{\infty} r
\int _{ \xi\in\mathcal A_m }
 \left(
\frac
{(\ln2)2^r\sigma^2}{P_m(d_{mi}(\xi))^{-\alpha}}
+\sum_{n\in\mathcal M\cup\{0\}\setminus \{m\}}
\frac{(\ln2)2^rP_n (d_{ni}(\xi))^{-\alpha}}
{(2^r-1) P_n (d_{ni}(\xi))^{-\alpha}+{P_m (d_{mi}(\xi))^{-\alpha}}}
\right)\right.\nonumber\\
&&\!\!\!\!\!\!\!\!\!\!\!\!\!\!
\quad\left.\exp\left(-\frac
{(2^r-1)\sigma^2}{P_m(d_{mi}(\xi))^{-\alpha}}
\right)
\prod_{n\in\mathcal M\cup\{0\}\setminus \{m\}}
\frac{P_m (d_{mi}(\xi))^{-\alpha}}{(2^r-1) P_n (d_{ni}(\xi))^{-\alpha}+{P_m (d_{mi}(\xi))^{-\alpha}}}
\lambda\:\text d
\xi \:\text d r
\right]^{-1},
\end{eqnarray}
where $\kappa_m=\frac { \text e^{-\lambda A_m} } { 1- \text e^{-\lambda A_m}  } (\text{Ei}(\lambda A_m) -\ln (\lambda A_m) - \gamma )$,
$A_m=\pi r_m^2$ for $m\in\mathcal M$, $A_0=\pi \left( r_0^2 - \sum_{ n\in \mathcal M } r_n^2 \right)$, $\text E \text i (x)=- \int_{\infty}^{-x} \frac {\text e^{t}}{t} \:\text dt$ is the exponential integral function \cite{gradshteyn2014table}, and $\gamma\approx 0.577$ is the Euler-Mascheroni constant \cite{weisstein2002euler}.
The average download time multiplied by the bandwidth per bit of pico BS
$m\in\mathcal M$  when downloading file from  the macro BS is given by
\begin{equation}\label{PFeq2}
b_m=\frac{-\ln2}
{\text e^{\frac{\sigma^2}{P_0   d_{m}^{-\alpha}}}
\text{Ei}\left(-{\frac{\sigma^2}{P_0   d_{m}^{-\alpha}}}\right)},\quad
\forall m \in \mathcal M.
\end{equation}
\end{lemma}

\itshape \textbf{Proof:}  \upshape
Please refer to Appendix A.
 \hfill $\Box$

\section{Optimal Cache Placement with Fixed Bandwidth Allocation}
Since the objective function (\ref{PFmin1}a) is nonconvex with respect to ($\pmb s, \pmb w$), it is in general hard to obtain the globally optimal solution of Problem (\ref{PFmin1}).
In this section, we investigate the optimization of cache placement with fixed bandwidth allocation.
Given bandwidth allocation, the original Problem (\ref{PFmin1}) can be decoupled into multiple cache placement problems for $M$ pico BSs, which fortunately have optimal solutions in closed form.

With given $\pmb w$,  Problem (\ref{PFmin1}) becomes the following problem:
\begin{subequations}\label{OCPmin1}
\begin{align}
\mathop{\min}_{ \pmb{s}}& \:\;
\sum_{m\in\mathcal M} \sum_{f\in\mathcal F}q_f\left(\frac{b_m(1-s_{mf})L_f}{w_m} +\frac{(1-s_{mf})L_fD}
{C_m-\sum_{f\in\mathcal F}s_{mf}L_f}
\right)
\\
\textrm{s.t.}
& \:\;\sum_{f\in\mathcal F}s_{mf}L_f\leq C_m, \quad \forall m \in \mathcal M\\
& \:\; 0\leq s_{mf}\leq 1,\quad\forall m\in\mathcal M, f \in \mathcal F.
\end{align}
\end{subequations}
Since Problem (\ref{OCPmin1}) has a decoupled objective function and decoupled constraints, Problem (\ref{OCPmin1}) can be decoupled into $M$ subproblems.
Subproblem $m$ for cache optimization in pico BS $m$ is formulated as
\begin{subequations}\label{OCPmin2}
\begin{align}
\mathop{\min}_{ \pmb{s}_m}& \:\;
\left(\frac{ b_m}{w_m} +\frac{D}
{C_m-\sum_{f\in\mathcal F}s_{mf}L_f}\right)\sum_{f\in\mathcal F}q_f L_f(1-s_{mf})
\\
\textrm{s.t.}
& \:\;\sum_{f\in\mathcal F}s_{mf}L_f\leq C_m\\
& \:\; 0\leq s_{mf}\leq 1,\quad\forall f \in \mathcal F,
\end{align}
\end{subequations}
where $\pmb s_m=[s_{m1},\cdots,s_{mF}]^T$.

Problem (\ref{OCPmin2}) is nonconvex.
To show this, we define function $g(x,y)=\frac{x}{C_m-L_f+x-y}$, $x,y\geq 0$, $y-x<C_m-L_f$.
The Hessian matrix of $g(x,y)$ is given by
\begin{eqnarray}
\bigtriangledown^2 g(x,y)
&&\!\!\!\!\!\!\!\!\!\!
=
\begin{pmatrix}
\frac{\partial ^2 g(x,y)}{\partial x^2}&\frac{\partial ^2 g(x,y)}{\partial x\partial y} \\
\frac{\partial ^2 g(x,y)}{\partial x \partial y}&\frac{\partial ^2 g(x,y)}{\partial y^2}
\end{pmatrix}
\nonumber\\
&&\!\!\!\!\!\!\!\!\!\!
=\frac1 {(C_m-L_f+x-y)^3}
\begin{pmatrix}
2(y-C_m+L_f)&C_m-L_f-x-y\\
C_m-L_f-x-y&2x
\end{pmatrix}.
\end{eqnarray}
Since 
\begin{eqnarray}
|\bigtriangledown^2 g(x,y)|
&&\!\!\!\!\!\!\!\!\!\!
=
-\frac1{(1-x-y)^3} (C_m-L_f +x-y)^2 < 0,
\end{eqnarray}
Hessian matrix $\bigtriangledown^2 g(x,y)$ is not positive semi-definite, i.e., function $g(x,y)$ is not a convex function with respect to $(x,y)$.
Denoting $x=L_f(1-s_{mf})$ and $y=\sum_{l\in\mathcal F\setminus\{f\}}s_{ml}L_l$, we can show that the objective function (\ref{OCPmin2}a) is not convex, i.e., Problem (\ref{OCPmin2}) is a nonconvex problem.

Although Problem (\ref{OCPmin2}) is nonconvex, the optimal cache placement in each pico BS can be obtained in closed form.
First, a special structure of the optimal solution is revealed by Theorem~1.
Second, this special structure shows that the optimal solution has a finite solution space, where each possibly optimal solution can be obtained according to Theorem 1.
Third, the optimal cache placement is one of finite candidate solutions with the best objective value as summarized in Theorem~2.
Finally, the number of candidate solutions for the optimal cache placement is greatly reduced through investigating the property of the candidate solutions according to Lemmas 2 and 3.
\begin{theorem}
In the optimal cache placement $\pmb s_m^*$ of Problem (\ref{OCPmin2}), at most one file $f$ has the optimal cache probability with the range $s_{mf}^*\in(0,1)$, and for all the other files, i.e., $s_{ml}^*\in\{0,1\}$, $\forall l\in\mathcal F\setminus\{f\}$.
Besides, the optimal cache placement $\pmb s_m^*$ satisfies $s_{m1}^*\geq s_{m2}^*\geq \cdots\geq s_{mF}^*$.
\end{theorem}

\itshape \textbf{Proof:}  \upshape
Please refer to Appendix B.
 \hfill $\Box$

According to Theorem 1, it is optimal to store the high-popularity file with high priority independent of the file length.
This is reasonable since caching files with high popularity can improve file hit probability, and the average fronthaul transmission delay and buffer delay can be reduced.
The special structure of the optimal solution indicated in Theorem 1 shows that the optimal solution of Problem (\ref{OCPmin2}) has a finite solution space.
By comparing all these possibly optimal solutions, the optimal cache placement of Problem (\ref{OCPmin2}) is given in the following theorem.
\begin{theorem}
The optimal $\pmb s_m^*$ of Problem (\ref{OCPmin2}) is one of the following $F_{m1}$ potential solutions with the highest objective value (\ref{OCPmin2}a): $(\pmb 1_{f-1}, s_{mf}^*, \pmb 0_{F-f})$, $f=1, \cdots, F_{m1}$, where $\pmb 1_{f-1}$ and $\pmb 0_{F-f}$ are defined in (\ref{appenBeq5}),
\begin{equation}\label{OCPeq0}
F_{m1} =
\begin{cases}
\min_{f\in\mathcal F, \sum_{l=1}^f L_l >C_m} f & \mbox{if there exits no $f\in\mathcal F$ satisfying $\sum_{l=1}^f L_l =C_m $}\\
\min_{f\in\mathcal F, \sum_{l=1}^f L_l =C_m} f  & \mbox{if there exits $f\in\mathcal F$ satisfying $\sum_{l=1}^f L_l =C_m $}
\end{cases},
\end{equation}
and
\begin{equation}\label{OCPeq0_1}
s_{mf}^*\!=\!
\begin{cases}
s_{mf}^{\max}&\!\! \mbox{if $C_m\geq \sum_{l=1}^{f}L_l+\sum_{l=f+1}^F\frac{q_lL_l}{q_f} $}\\
\arg\min_{s_{mf}\in\{s_{mf}(1),s_{mf}^{\max}\}}g_{mf}(s_{mf}) & \!\!\mbox{if $C_m <  \sum_{l=1}^{f}L_l + \sum_{l=f+1}^F\frac{q_lL_l}{q_f}  $
and $0\!<\!s_{mf}(1)\!<\!1$}\\
\arg\min_{s_{mf}\in\{0,s_{mf}^{\max}\}}g_{mf}(s_{mf}) &\!\!
\mbox{if $C_m< \sum_{l=1}^{f}L_l+\sum_{l=f+1}^F\frac{q_lL_l}{q_f} $ and $s_{mf}(1)\leq 0$}
\\
s_{mf}^{\max} &\!\!
\mbox{if $C_m< \sum_{l=1}^{f}L_l+\sum_{l=f+1}^F\frac{q_lL_l}{q_f}$ and $s_{mf}(1)>s_{mf}^{\max}$}
\end{cases}\!\!,
\end{equation}

with $s_{mf}^{\max}$, $g_{mf}(s_{mf})$ and $s_{mf}(1)$ defined in (\ref{apenCeq0}), (\ref{appenCeq1}a) and (\ref{appenCeq2_5}), respectively.
\end{theorem}

\itshape \textbf{Proof:}  \upshape
Please refer to Appendix C.
 \hfill $\Box$

Theorem 2 shows the impact of file popularity, file length and storage capacity on the optimal caching strategy.
Furthermore, Theorem 2 points out that the optimal solution is one of the $F_{m1}$ potential solutions.
Based on Theorem 2, we can infer the following two lemmas.

\begin{lemma}
For two solutions $(\pmb 1_{f-1},  \pmb 0_{F-f+1})$ and $(\pmb 1_{f-1}, s_{mf}^*, \pmb 0_{F-f})$ with $s_{mf}^*\in(0,1]$, solution $(\pmb 1_{f-1}, s_{mf}^*, \pmb 0_{F-f})$ yields better objective value (\ref{OCPmin2}a) than solution $(\pmb 1_{f-1},  \pmb 0_{F-f+1})$.
\end{lemma}

\itshape \textbf{Proof:}  \upshape
According to the proof of Theorem 2, $s_{mf}^*$ is the optimal cache strategy with given $s_{ml}=1$ for $l<f$ and $s_{ml}=0$ for $l>f$, which shows that the objective function (\ref{OCPmin2}a) of solution $(\pmb 1_{f-1}, s_{mf}^*, \pmb 0_{F-f})$ is smaller than that of solution  $(\pmb 1_{f}, \pmb 0_{F-f})$, i.e., Lemma 2 is proved.

 \hfill $\Box$

\begin{lemma}
The optimal $\pmb s_m^*$ of Problem (\ref{OCPmin2}) is one of the following $F_{m1}-F_{m2}+1$ potential solutions with the highest objective value (\ref{OCPmin2}a): $(\pmb 1_{f-1}, s_{mf}^*, \pmb 0_{F-f})$, $f=F_{m2}, \cdots, F_{m1}$, where
\begin{equation}\label{OCPeq0_2}
F_{m2}=\max_{f\in\mathcal F, C_m\geq \sum_{l=1}^f L_l +\sum_{l=f+1}^F\frac{q_lL_l}{q_f}} f,
\end{equation}
and $s_{mf}^*$ is given by (\ref{OCPeq0_1}).
\end{lemma}

\itshape \textbf{Proof:}  \upshape
Denoting $z(f)=\sum_{l=1}^fL_l+\sum_{l=f+1}^F\frac{q_lL_l}{q_f}$, we have
\begin{align}\label{OCPeq0_3}
z(f)-z(f-1)&=L_f+\sum_{l=f+1}^F\frac{q_l L_l}{q_f}-\sum_{l=f}^F\frac{q_l L_l}{q_{f-1}}
>L_f+\sum_{l=f+1}^F\frac{q_l L_l}{q_f}-\sum_{l=f}^F\frac{q_l L_l}{q_{f}}
=0,
\end{align}
where the inequality follows from that $q_f<q_{f-1}$.
Based on (\ref{OCPeq0_2}) and (\ref{OCPeq0_3}), we have $C_m\geq z(F_{m2})> z(F_{m2}-1)>\cdots>z(1)$. 
From Theorem 2, the first $F_{m2}$ potentially optimal solutions can be expressed by $(\pmb 1_{f-1}, s_{mf}^{\max}, \pmb 0_{F-f})$, $f=1, \cdots, F_{m2}$.
Since (\ref{OCPeq0_2}) implies that $\frac{C_m-\sum_{l=1}^{f-1}L_l}{L_f}\geq1$, we have $s_{mf}^{\max}=1$ for $f\leq F_{m2}$ according to (\ref{apenCeq0}).
As a result, the first $F_{m2}$ potentially optimal solutions are $(\pmb 1_{f-1}, 1, \pmb 0_{F-f})$, $f=1, \cdots, F_{m2}$.
Combining Theorem 2 and Lemma 2, Lemma 3 is proved.
 \hfill $\Box$

According to Lemma 3, the number of potentially optimal solutions can be reduced to $F_{m1}-F_{m2}+1$, which can largely simplify the computation of obtaining the optimal solution.

\section{Joint Bandwidth and Cache Optimization}
In this section, we jointly optimize bandwidth allocation and cache placement to solve Problem (\ref{PFmin1}).
The optimal bandwidth allocation can be obtained in closed from by checking the KKT conditions.
Based on the result of the optimal bandwidth allocation, the original Problem (\ref{PFmin1}) is equivalent to a problem with only cache placement variables.
To solve the equivalent nonconvex problem, we derive one suboptimal algorithm with low complexity.

\subsection{Optimal Bandwidth Allocation}
With given $\pmb s$, Problem (\ref{PFmin1}) becomes the following problem.
\begin{subequations}\label{OBAmin1}
\begin{align}
\mathop{\min}_{\pmb w}& \:\;
 \frac{ \sum_{f\in\mathcal F}q_f L_f \sum_{m\in\mathcal M\cup\{0\}}a_m}{w_0}+
\sum_{m\in\mathcal M}
\frac{\sum_{f\in\mathcal F} b_m q_f L_f(1-s_{mf})}{w_m}
\\
\textrm{s.t.}
& \:\; \sum_{m\in\mathcal M\cup\{0\}} w_m \leq W\\
& \:\; w_m\geq0,\quad\forall m\in\mathcal M\cup\{0\}.
\end{align}
\end{subequations}
Since the objective function (\ref{OBAmin1}a) is convex and the constraints (\ref{OBAmin1}b)-(\ref{OBAmin1}c) are all linear,
Problem (\ref{OBAmin1}) is a convex problem, which can be globally optimal solved via the KKT conditions \cite{Yang2017Joint,xu2010mimo,Yang2018EEIoT,Yang2018Optimal}.
Thus, the following theorem is provided.
\begin{theorem}
The optimal bandwidth allocation to Problem (\ref{OBAmin1}) is
\begin{equation}\label{OBAeq1}
w_0=\frac{ aW}
{ a +\sum_{m\in\mathcal M}\sqrt{\sum_{f\in\mathcal F}b_m q_f L_f (1-s_{mf})}},
\end{equation}
and
\begin{equation}\label{OBAeq1_2}
w_m=\frac{\sqrt{\sum_{f\in\mathcal F} b_m q_f L_f(1-s_{mf})}W}
{a +\sum_{m\in\mathcal M}\sqrt{\sum_{f\in\mathcal F}b_mq_f L_f  (1-s_{mf})}}, \quad \forall m \in \mathcal M,
\end{equation}
where $a=\sqrt{\sum_{f\in\mathcal F}q_fL_f\sum_{m\in\mathcal M\cup\{0\}}a_m}$.
\end{theorem}

\itshape \textbf{Proof:}  \upshape
Please refer to Appendix D.
 \hfill $\Box$

\subsection{Cache Optimization}

Substituting the optimal bandwidth allocation (\ref{OBAeq1}) and (\ref{OBAeq1_2}) into Problem (\ref{PFmin1}) yields
\begin{subequations}\label{JBCmin1}
\begin{align}
\mathop{\min}_{ \pmb{s} }& \:\;
\frac1W \left( a +\sum_{m\in\mathcal M}\sqrt{\sum_{f\in\mathcal F}b_m q_f L_f (1-s_{mf})}\right)^2
+\sum_{m\in\mathcal M}
\frac{ \sum_{f\in\mathcal F}q_fL_fD(1-s_{mf})}
{C_m-\sum_{f\in\mathcal F}L_f s_{mf}}
\\
\textrm{s.t.}
& \:\;\sum_{f\in\mathcal F}s_{mf}L_f\leq C_m, \quad \forall m \in \mathcal M\\
& \:\; 0\leq s_{mf}\leq 1,\quad\forall m\in\mathcal M, f \in \mathcal F.
\end{align}
\end{subequations}

Due to the nonconvex objective function (\ref{JBCmin1}a), Problem (\ref{JBCmin1}) is a nonconvex problem.
To solve nonconvex Problem (\ref{JBCmin1}), we provide the following theorem to exploit the optimal structure of the optimal solution.

\begin{theorem}
In the optimal cache placement $\pmb s^*$ to Problem (\ref{JBCmin1}), for each pico BS $m\in\mathcal M$, at most one files $f_m$ has the optimal cache probability with the range $s_{mf_m}^*\in(0,1)$, and for all the other files, i.e., $s_{ml}^*\in\{0,1\}$, $\forall l\in\mathcal F\setminus\{f\}$.
Besides, the optimal cache placement $\pmb s^*$ satisfies $s_{m1}^*\geq s_{m2}^*\geq \cdots\geq s_{mF}^*$, for all $m \in \mathcal M$.
\end{theorem}

\itshape \textbf{Proof:}  \upshape
Please refer to Appendix E.
 \hfill $\Box$

According to Theorem 4, it is optimal to store the high-popularity files with high priority as indicated from Theorem 1.
Theorem 4 reflects the structure of the optimal solution of Problem (\ref{JBCmin1}).
Since the cache strategies for different pico BSs are coupled in the objective function (\ref{JBCmin1}a), it is hard to obtain the globally optimal solution of nonconvex Problem (\ref{JBCmin1}).
In the following, we propose an iterative cache placement (ICP) algorithm to solve Problem (\ref{JBCmin1}) in Algorithm 1.

\begin{algorithm}[h]
\caption{Iterative Cache Placement (ICP)}
\label{alg:Framwork1}
\begin{algorithmic}[1]
\State Initialize $\pmb s_1^{(0)}, \cdots, \pmb s_M^{(0)}$.
Set $k=1$, and maximal iteration number $K_{\max}$.
\For{$m=1, 2, \cdots, M$}
\State Calculate the optimal $\pmb s_{m}^{(k)}$ to Problem (\ref{JBCmin1}) with given $(\pmb s_1^{(k)}, \cdots, \pmb s_{m-1}^{(k)}, \pmb s_{m+1}^{(k-1)}, \cdots, \pmb s_{M}^{(k-1)})$.
\EndFor
\State
If $k > K_{\max}$ or the objective function (\ref{JBCmin1}a) converges,
output $\pmb s^*=(\pmb s_1^{(k)}, \cdots, \pmb s_M^{(k)})$, and
terminate.
Otherwise,
set $k=k+1$ and go to step 2.
\end{algorithmic}
\end{algorithm}

In the ICP algorithm, to calculate the optimal $\pmb s_{m}$ to Problem (\ref{JBCmin1}) with given cache strategies of other pico BSs, we provide the following theorem.

\begin{theorem}
The optimal cache placement $\pmb s_m^*=[s_{m1}^*,\cdots,s_{mF}^*]$ of Problem (\ref{JBCmin1}) with given $[s_{11},\cdots,s_{(m-1)F},s_{(m+1)1},\cdots,s_{MF}]^T$ is one of the following $F_{m1}$ potential solutions with the highest objective value (\ref{JBCmin1}a): $(\pmb 1_{f-1}, s_{mf}^*, \pmb 0_{F-f})$, $f=1, \cdots, F_{m1}$,
$F_{m1}$ is defined in (\ref{OCPeq0}),
and
\begin{equation}\label{JBCeq1}
s_{mf}^*=
\arg\min_{s_{mf}\in\left\{0,s_{mf}^{\max},s_{mf}(1),
\cdots,s_{mf}(k)\right\}}y_{mf}(s_{mf}),
\end{equation}
where $s_{mf}^{\max}$ is defined in (\ref{apenCeq0}), $k\in\{0,1,2,3,4,5\}$, $s_{mf}(i)=\frac{b_m\sum_{l=f}^Fq_l L_l-(x(i))^2}{b_m q_f l_f}$, $i=1, \cdots, k$, and $x(1), \cdots, x(k)$ are $k$ roots in interval $(\sqrt{b_m \sum_{l=f} ^F q_lL_l-b_mq_fL_fs_{mf}^{\max}},$ $\sqrt{b_m \sum_{l=f} ^F q_lL_l})$ to equation (\ref{appenEeq2_2}).
\end{theorem}

\itshape \textbf{Proof:}  \upshape
Please refer to Appendix F.
 \hfill $\Box$

Theorem 5 indicates that the optimal $\pmb s^*_m$ has a finite solution space with $F_{m1}$ potential solutions, which ensures that the optimal cache placement for each pico BS can be effectively obtained by comparing finite solutions.

\subsection{Complexity Analysis}

For the ICP algorithm, the major complexity in each iteration lies in  calculating the optimal $\pmb s_{m}^{(k)}$ to Problem (\ref{JBCmin1}) with given cache placement of other pico BSs.
The optimal $\pmb s_{m}^{(k)}$ is one of the $F_{m1}$ potential solutions according to Theorem 5.
The complexity of obtaining each potential solution is $\mathcal O(T)$, where $T$ is the complexity of calculating the roots  to equation (\ref{appenEeq2_2}) via root-finding algorithm for polynomials \cite{madsen1973root,xu2011joint,xu2010mimo}.
As a result, the total complexity of the ICP algorithm is $\mathcal O(KT\sum_{m\in\mathcal M}F_{m1})$, where $K$ is the number of iterations of the ICP algorithm.

\section{Numerical Results}

In this section, numerical results are presented to evaluate the performance of the proposed ICP algorithm.
In the simulations, we consider a circular macrocell with three pico BSs, i.e., $M=3$.
The macro cell has radius $r_0=1$ km, and the coverage area of each pico BS is a circle area with radius $r_1=\cdots=r_M=150$ m.
The macro BS is located at the origin, and the pico BSs are located at $(-339,741)$, $(218,-230)$, $(561,-457)$.
The path-loss exponent is set as $\alpha=3.76$, and the distribution of users is modeled as independent PPP of density $\lambda=500$ /km$^2$.
We assume equal file length, i.e., $L_1=\cdots=L_F=L$, and equal storage capacity for each pico BS, i.e., $C_1=\cdots=C_M=C$.
The number of files is $F=1000$, and the Zipf distribution is adopted to model the content popularity distribution \cite{4801529}
\begin{equation}\label{numeq1}
q_f=\frac{1/f^{\nu}}{\sum_{l=1}^F 1/l^{\nu}}, \quad \forall f \in \mathcal F,
\end{equation}
where $\nu\geq0$ stands for the skewness of popularity distribution, and larger $\nu$ represents more centralized file request.
Unless specified otherwise, system parameters are set as $\nu=0.8$, $W=10$ MHz, $D=5$ s, and $C=1000$ Mbits.

We compare the proposed ICP algorithm with the optimal cache placement with equal bandwidth allocation (labeled as `OCEB') algorithm, where the optimal cache placement is obtained from Theorem 2 and $w_0=w_1=\cdots=w_M=\frac{W}{M+1}$, and the optimal cache placement algorithm \cite{liu2017optimizing} with fixed buffer capacity (half of the storage capacity is left for buffering) and optimized bandwidth allocation obtained from the ICP (labeled as `OCFBOB').

\begin{figure}
\centering
\includegraphics[width=3.6in]{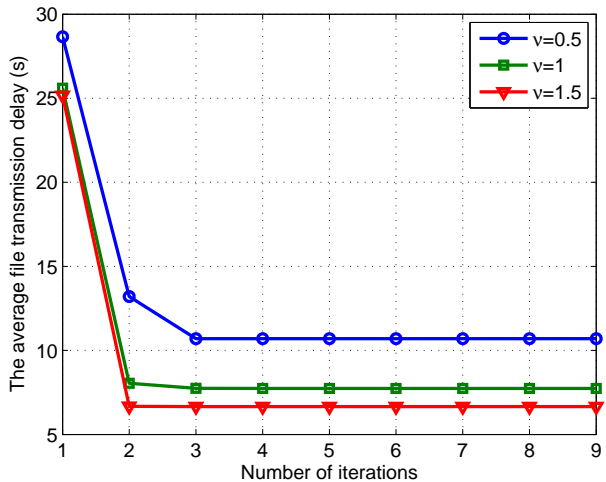}
\caption{Convergence behavior of the ICP algorithm under different values of parameter $\nu$.}\label{numfig3}
\end{figure}

The convergence behavior of the ICP algorithm is illustrated in Fig.~\ref{numfig3}.
From this figure, the average file transmission delay monotonically decreases and converges rapidly.
Note that only two or three iterations are sufficient for the algorithm to converge, which shows the effectiveness of the proposed algorithm.

\begin{figure}
\centering
\includegraphics[width=3.6in]{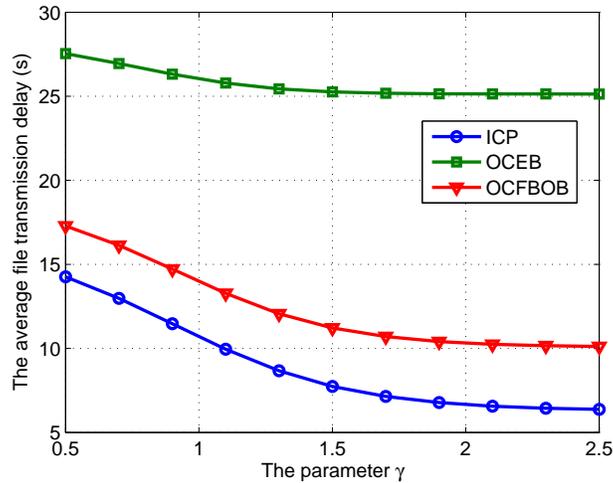}
\caption{The average file transmission delay versus the parameter $\nu$.}\label{numfig5}
\end{figure}

In Fig.~\ref{numfig5}, we show the average file transmission delay versus different parameters $\nu$ in (\ref{numeq1}).
From Fig.~\ref{numfig5}, it is observed that the ICP outperforms the other two algorithms in terms of delay.
This is because the ICP jointly optimizes bandwidth allocation and cache placement, while the OCEB only optimizes cache placement and the OCFBEB ignores the tradeoff of cache and buffer.
The OCEB yields the largest delay among three algorithms, which shows that the optimization of bandwidth allocation can greatly reduce the delay.

\begin{figure}
\centering
\includegraphics[width=3.6in]{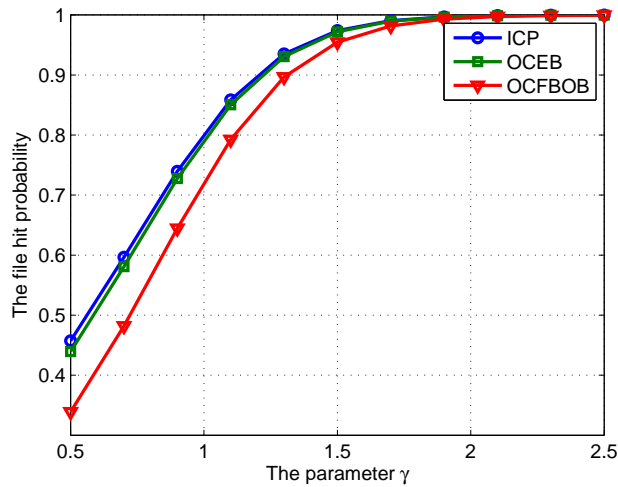}
\caption{The file hit probability versus the parameter $\nu$.}\label{numfig6}
\end{figure}

The file hit ratio versus different parameters $\nu$ is depicted in Fig.~\ref{numfig6},
where the file hit ratio $\rho$ is defined as the average successful probability that a user can fetch files from the pico BSs, i.e.,
$\rho=\sum_{m\in\mathcal M} \frac 1 M \sum_{f\in\mathcal F} q_f s_{mf}$.
According to Fig.~\ref{numfig6}, the file hit probabilities of the ICP and OCEB are larger than that of the OCFBEB.
This is because that the ICP and OCEB consider the tradeoff of cache and buffer to cache more files by reducing the buffer capacity, while the OCFBEB assumes fixed buffer capacity and the cache capacity cannot be further improved.
From Fig.~\ref{numfig6}, we find that the file hit probability of the ICP is slightly larger than that of the OCEB, while the average delay of the ICP is significantly superior over that of the OCEB according to Fig.~\ref{numfig5}.
This is because that the delay performance not only depends on the users' hit performance, but also on the transmission rate and buffer delay.
Combining Fig.~\ref{numfig5} and Fig.~\ref{numfig6}, we can conclude that the ICP achieves the best delay performance as well as the highest file hit probability through considering two tradeoffs: cache placement versus bandwidth allocation, and cache capacity versus buffer capacity, i.e., the ICP outperforms the OCFBEB in terms of delay through increasing file hit probability, and the ICP outperforms the OCEB in terms of delay through proper bandwidth allocation.

\begin{figure}
\centering
\includegraphics[width=3.6in]{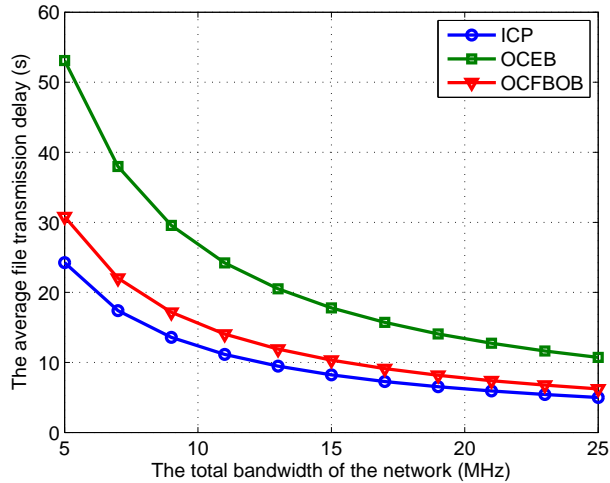}
\caption{The average file transmission delay versus the total system bandwidth.}\label{numfig7}
\end{figure}

As shown in Fig.~\ref{numfig7}, we illustrate the average file transmission delay versus the total system bandwidth $W$.
It is found that the average file transmission delay decreases with the total system bandwidth, since large bandwidth leads to large file transmission rate.
It is also observed that the ICP outperforms the other two algorithms, and the delay is greatly reduced by using  the ICP compared to the OCEB when the total system bandwidth is small.
This is due to the fact that the bandwidth is optimally allocated in the ICP, which results in good performance especially for limited system bandwidth resource.

\begin{figure}
\centering
\includegraphics[width=3.6in]{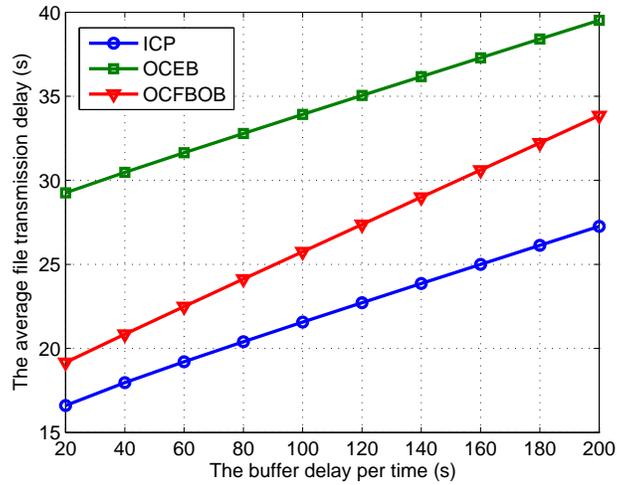}
\caption{The average file transmission delay versus the buffer delay per time.}\label{numfig8}
\end{figure}

Fig.~\ref{numfig8} demonstrates the average file transmission delay versus the buffer delay $D$ per time.
It can be seen that  the average file transmission delay increases with the buffer delay for all algorithms.
It is also found that the ICP yields the best performance in terms of  delay, and the delay is greatly improved by using the ICP compared to the OCFBEB for large buffer delay.
The reason is that the ICP can dynamically allocate cache capacity and buffer capacity to reduce the delay based on the value of buffer delay, while the OCFBEB assumes fixed cache capacity and buffer capacity allocation.

\begin{figure}
\centering
\includegraphics[width=3.6in]{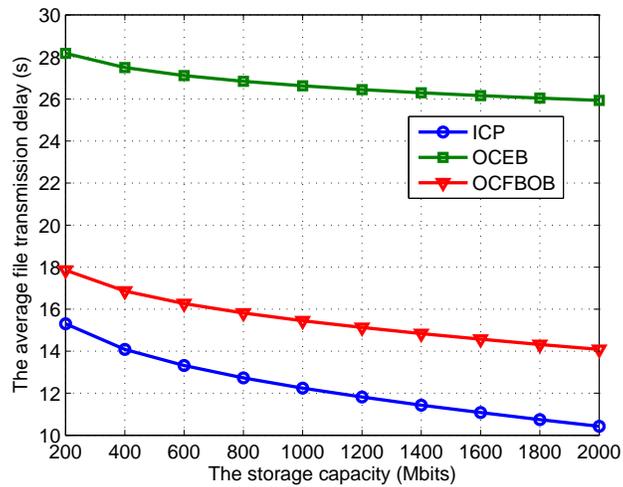}
\caption{The average file transmission delay versus the storage capacity.}\label{numfig9}
\end{figure}

\begin{figure}
\centering
\includegraphics[width=3.6in]{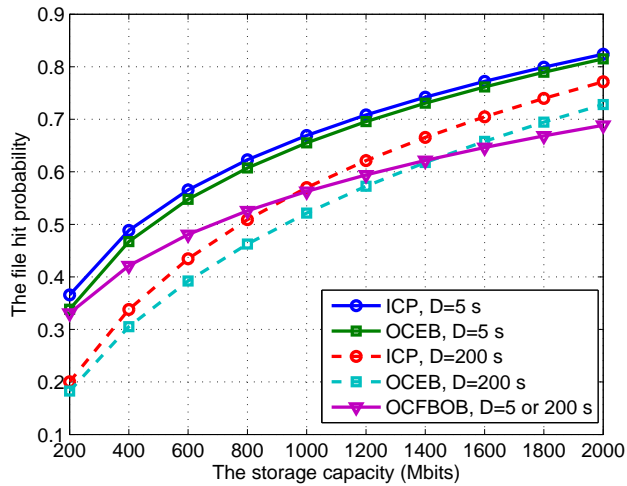}
\caption{The file hit probability versus the storage capacity} \label {numfig10}
\end{figure}

In Fig.~\ref{numfig9} and Fig.~\ref{numfig10}, we show the average file transmission delay and file hit probability versus different storage capacities $C$, respectively.
According to Fig.~\ref{numfig9}, the average file transmission delay monotonically decreases with the increase of the storage capacity.
The is due to the following two reasons.
One reason is that with the increase of storage capacity, more files can be cached in the pico BSs and more users can download files directly from the cache of the pico BSs.
The other reason is that larger storage capacity can lead to larger buffer capacity, which reduces the buffer time.
Fig.~\ref{numfig10} illustrates that the file hit probability increases with the storage capacity, since the pico BSs can cache more popular files for larger storage capacity.
For small buffer delay $D$, the file hit probability of the ICP or OCEB is larger than that of the OCFBEB.
This is because the average delay mainly lies in the transmission delay for small $D$ and large cache capacity is allocated by the ICP and OCEB.
For large buffer delay $D$ and small storage capacity $C$, the file hit probability of the OCFBEB is superior over that of the ICP and OCEB.
This is due to the fact that the buffer delay consumption dominates the transmission delay for large $D$ and limited $C$, which allows to allocate more capacity to buffer by the ICP and OCEB.


\section{Conclusions}

In this paper, we investigated the tradeoff of cache capacity and buffer capacity via joint bandwidth allocation and cache placement to minimize the average file transmission delay.
By analyzing the KKT conditions of the nonconvex delay minimization problem, we show that it is optimal to cache the files with high popularity first rather than the files with large size.
We proposed an iterative algorithm to obtain a suboptimal solution with low complexity.
Through dynamically allocating cache capacity and buffer capacity, the proposed algorithm is superior over the existing caching strategy with fixed buffer capacity.
It tends to allocate more cache capacity for low buffer delay per time and high storage capacity, while more capacity should be shifted to the buffer capacity for high buffer delay per time and low storage capacity.

\appendices
\section{Proof of Lemma 1}
\setcounter{equation}{0}
\renewcommand{\theequation}{\thesection.\arabic{equation}}
According to (\ref{sys2eq1}) and (\ref{sys2eq2}), we have
\begin{equation}\label{appen0Eq1}
a_{m}={\frac {w_0}{\mathbb E_{U_m,h_{0i},\cdots,h_{Mi},\xi}R_{mi}(\xi)}}
= \frac{1}{\mathbb E_{U_m}\frac 1{U_m}\mathbb E_{h_{0i},\cdots,h_{Mi},\xi} \bar R_{mi}(\xi)},
\end{equation}
for all $m\in\mathcal M\cup\{0\}$, where
$\bar R_{mi}(\xi)=\log_2\left(1+
\frac{P_m |h_{mi}|^2 (d_{mi}(\xi))^{-\alpha}}
{\sigma^2+\sum_{n\in\mathcal M\cup\{0\}\setminus \{m\}}P_n |h_{ni}|^2 (d_{ni}(\xi))^{-\alpha}}\right)$.
Assume that $U_m>0$, $\forall m \in \mathcal M\cup\{0\}$.
Since users follow independent PPP with density $\lambda$,
we have
\begin{equation}\label{appen0Eq1_1}
\mathbb P(U_m=k)= \frac 1 { 1- \text e^{-\lambda A_m}  }
\text e^{-\lambda A_m}\frac{(\lambda A_m)^k}{k!},\quad k=1, 2, \cdots
\end{equation}
for all $m \in \mathcal M\cup\{0\}$, where $A_m=\pi r_m^2$ for all $m\in\mathcal M$, $A_0=\pi r_0^2-\sum_{m\in\mathcal M} r_m^2$,
and $\frac{1}{ 1- \text e^{-\lambda A_m}  }$ is the modified parameter to ensure that $\sum_{k=1}^{\infty}\mathbb P(U_m=k)=1$.
Based on (\ref{appen0Eq1_1}), we can obtain
\begin{equation}\label{appen0Eq1_2_1}
\mathbb E_{U_m} \frac 1 {U_m} = \frac { \text e^{-\lambda A_m} } { 1- \text e^{-\lambda A_m}  } \sum_{k=1}^ {\infty} \frac{(\lambda A_m)^k}{kk!}.
\end{equation}
Define $z(\lambda)= \sum_{k=1}^ {\infty} \frac{(\lambda A_m)^k}{kk!}$, and the derivative of $z(\lambda)$ is
\begin{equation}\label{appen0Eq1_2_2}
z'(\lambda)=\sum_{k=1}^ {\infty} \frac{A_m(\lambda A_m)^{(k-1)}}{k!}=\frac 1{\lambda }
\sum_{k=1}^ {\infty} \frac{(\lambda A_m)^{k}}{k!}
=\frac1{\lambda} (\text e^{\lambda A_m}-1),
\end{equation}
where the last equality follows from the fact that $\text e^{\lambda A_m}=\sum_{k=0}^ {\infty} \frac{(\lambda A_m)^{k}}{k!}$.
Integrating both sides of equation (\ref{appen0Eq1_2_2}) yields
\begin{equation}\label{appen0Eq1_2_3}
z(\lambda)=\text{Ei}(\lambda A_m) -\ln (\lambda A_m) +C_{\text{cont}},
\end{equation}
which follows from the definition of the exponential integral function $\text{Ei}(\cdot)$ \cite[Equation 2.325]{gradshteyn2014table}.
In (\ref{appen0Eq1_2_2}), $C_{\text{cons}}$ is a constant.
Since $z(0)=0$, $C_{\text{cont}}=\lim_{\lambda\rightarrow0}\ln (\lambda A_m)-\text{Ei}(\lambda A_m)=-\gamma$ according to \cite[Page 252]{bender2013advanced}, where $\gamma$ is the Euler-Mascheroni constant.
Based on (\ref{appen0Eq1_2_1}) an (\ref{appen0Eq1_2_2}), we have
\begin{equation}\label{appen0Eq1_2}
\mathbb E_{U_m} \frac 1 {U_m} = \frac { \text e^{-\lambda A_m} } { 1- \text e^{-\lambda A_m}  } ( \text{Ei} (\lambda A_m) -\ln (\lambda A_m) - \gamma ).
\end{equation}

To calculate ${\mathbb E_{h_{0i},\cdots,h_{Mi},\xi}\bar R_{mi}(\xi)}$, we first calculate the complementary cumulative distribution function (CCDF):
\begin{eqnarray}\label{appen0Eq1_3}
&&\!\!\!\!\!\!\!\!\!\!\!\!\quad \mathbb P[\bar R_{mi}(\xi)>r]
 =
\mathbb P\left[ |h_{mi}|^2
>\frac
{(2^r-1)\left(\sigma^2+\sum_{n\in\mathcal M\cup\{0\}\setminus \{m\}}P_n |h_{ni}|^2 (d_{ni}(\xi))^{-\alpha}\right)}
{P_m (d_{mi}(\xi))^{-\alpha}}  \right]
\nonumber\\
&&\!\!\!\!\!\!\!\!\!\!\!\!\overset{(\text a)}{=} \int _{ \xi\in\mathcal A_m }
\mathbb E_{h_{ni},{\forall n\in\mathcal M\cup\{0\}\setminus \{m\}}}
\exp\left(-\frac
{(2^r-1)\left(\sigma^2+\sum_{n\in\mathcal M\cup\{0\}\setminus \{m\}}P_n |h_{ni}|^2 (d_{ni}(\xi))^{-\alpha}\right)}
{P_m (d_{mi}(\xi))^{-\alpha}}
\right)
\lambda\:\text d \xi
\nonumber\\
&&\!\!\!\!\!\!\!\!\!\!\!\!=\int _{ \xi\in\mathcal A_m }
\exp\left(-\frac
{(2^r-1)\sigma^2}{P_m (d_{mi}(\xi))^{-\alpha}}
\right)
\prod_{n\in\mathcal M\cup\{0\}\setminus \{m\}} \mathcal L_{|h_{ni}|^2}
\left(\frac{(2^r-1) P_n (d_{ni}(\xi))^{-\alpha}}{P_m (d_{mi}(\xi))^{-\alpha}}
\right)
\lambda\:\text d
\xi
\nonumber\\
&&\!\!\!\!\!\!\!\!\!\!\!\!
\overset{(\text b)} =\!\int _{ \xi\in\mathcal A_m }
\exp\left(\!-\!\frac
{(2^r-1)\sigma^2}{P_m(d_{mi}(\xi))^{-\alpha}}
\!\right)\!\!
\!\prod_{n\in\mathcal M\cup\{0\}\setminus \{m\}}
\frac{P_m (d_{mi}(\xi))^{-\alpha}}{(2^r-1) P_n (d_{ni}(\xi))^{-\alpha}+{P_m (d_{mi}(\xi))^{-\alpha}}}
\lambda\:\text d
\xi,
\end{eqnarray}
where both (a) and (b) follow from that $|h_{ni}|^2 \sim \exp(1)$, $\forall n\in\mathcal M\cup\{0\}$,
$\mathcal L_{|h_{ni}|^2}(\cdot)$ is the Laplace transform of $|h_{ni}|^2$.
Based on (\ref{appen0Eq1_3}), the probability density function (PDF) of $\bar R_{mi}(\xi)$ is
\begin{eqnarray}\label{appen0Eq2}
f_{\bar R_{mi}(\xi)} (r)&&\!\!\!\!\!\!\!\!\!\!=
\int _{ \xi\in\mathcal A_m }
 \left(
\frac
{(\ln2)2^r\sigma^2}{P_m(d_{mi}(\xi))^{-\alpha}}
+\sum_{n\in\mathcal M\cup\{0\}\setminus \{m\}}
\frac{(\ln2)2^rP_n (d_{ni}(\xi))^{-\alpha}}
{(2^r-1) P_n (d_{ni}(\xi))^{-\alpha}+{P_m (d_{mi}(\xi))^{-\alpha}}}
\right)\nonumber\\
&&\!\!\!\!\!\!\!\!\!\!\!\!\!\!\!\!
\exp\left(-\frac
{(2^r-1)\sigma^2}{P_m(d_{mi}(\xi))^{-\alpha}}
\right)
\prod_{n\in\mathcal M\cup\{0\}\setminus \{m\}}
\frac{P_m (d_{mi}(\xi))^{-\alpha}}{(2^r-1) P_n (d_{ni}(\xi))^{-\alpha}+
{P_m (d_{mi}(\xi))^{-\alpha}}}
\lambda\:\text d
\xi.
\end{eqnarray}
Combining (\ref{appen0Eq1}), (\ref{appen0Eq1_2}) and (\ref{appen0Eq2}), $a_m$
can be expressed as (\ref{PFeq1}).

From (\ref{sys2eq3}), we can obtain
\begin{eqnarray}\label{appen0Eq3}
b_m=\frac{w_m}{ \mathbb E_{ h_m } R_m }
=\frac 1 {\bar R_m},
\end{eqnarray}
where
\begin{eqnarray}\label{appen0Eq3_2}
\bar R_m&&\!\!\!\!\!\!\!\!\!={ \mathbb E_{ h_m } \log_2\left(1+
\frac{P_0 |h_{m}|^2 d_{m}^{-\alpha}}
{\sigma^2}\right)}
\nonumber\\
&&\!\!\!\!\!\!\!\!\!
=
\int_0^{\infty} \log_2\left(1+
\frac{P_0   d_{m}^{-\alpha}}
{\sigma^2}x\right) \text e^{-x} \: \text dx
\nonumber\\
&&\!\!\!\!\!\!\!\!\!\overset{(\text c)} =
-\left.\log_2\left(1+
\frac{P_0   d_{m}^{-\alpha}}
{\sigma^2}x\right)\text e^{-x}\right|_0^{\infty}
+\int_0^{\infty}
\frac{ \text e^{-x} }
{(\ln2)( x+\sigma^2/(P_0   d_{m}^{-\alpha}))} \: \text dx
\nonumber\\
&&\!\!\!\!\!\!\!\!\!\overset{(\text d)} =
-\frac{1}
{\ln2} \text e^{\frac{\sigma^2}{P_0   d_{m}^{-\alpha}}}
\text{Ei}\left(-{\frac{\sigma^2}{P_0   d_{m}^{-\alpha}}}\right),
\end{eqnarray}
where we obtain (c) by using integration by parts and (d)
follows from \cite[Equation (3.352.4)]{gradshteyn2014table}.
Based on (\ref{appen0Eq3}) and (\ref{appen0Eq3_2}), $b_m$ is given by
(\ref{PFeq2}).
\section{Proof of Theorem 1}
\setcounter{equation}{0}
\renewcommand{\theequation}{\thesection.\arabic{equation}}

Introducing auxiliary variable $t_m$, Problem (\ref{OCPmin2}) is equivalent to
\begin{subequations}\label{appenBeq1}
\begin{align}
\mathop{\min}_{ \pmb{s}_m, t_m}& \:\;
\left(\frac{ b_m}{w_m} +\frac{D}{C_m-t_m}\right)\sum_{f\in\mathcal F}q_f L_f(1-s_{mf})
\\
\textrm{s.t.}
& \:\; \sum_{f\in\mathcal F}s_{mf} L_f \leq t_m \\
& \:\; t_m\leq C_m\\
& \:\; 0\leq s_{mf}\leq 1,\quad\forall f \in \mathcal F,
\end{align}
\end{subequations}
since constraint (\ref{appenBeq1}b) always holds with equality for the optimal solution.
With given $t_m$, the Lagrangian function of Problem (\ref{appenBeq1}) is
\begin{eqnarray}\label{apenBeq2}
\mathcal L_1(\pmb s_m, \theta_m, \pmb \phi_m, \pmb \psi_m)
\!\!\!&=&\!\!\!
\left(\frac{ b_m}{w_m} +\frac{D}{C_m-t_m}\right)\sum_{f\in\mathcal F}q_f L_f(1-s_{mf})
+\theta_m\left( \sum_{f\in\mathcal F} s_{mf}L_f-t_m\right)
\nonumber\\
&&\!\!\!+\sum_{f\in\mathcal F}\phi_{mf}(-s_{mf})+\sum_{f\in\mathcal F}\psi_{mf}(s_{mf}-1),
\end{eqnarray}
where $\theta_m$, $\pmb \phi_m=[\phi_{m1},\cdots,\phi_{mF}]^T$ and $\pmb \psi_m=[\psi_{m1},\cdots,\psi_{mF}]^T$ are nonnegative Lagrangian multipliers associated with corresponding  constraints of Problem (\ref{appenBeq1}).
According to \cite{boyd2004convex,Yang2018Association,Yang2017On,Yang2017Energy}, the optimal solution should satisfy the following KKT conditions of Problem (\ref{appenBeq1}):
\begin{equation}\label{appenBKKT1}
\frac {\partial \mathcal L_1}{\partial s_{mf}}=
-\left(\frac{ b_m}{w_m} +\frac{D}{C_m-t_m}\right)q_fL_f+\theta_m L_f-\phi_{mf}+\psi_{mf}=0, \quad \forall f\in\mathcal F.
\end{equation}
From (\ref{appenBKKT1}), we have
\begin{equation}\label{appenBeq3}
-\left(\frac{ b_m}{w_m} +\frac{D}{C_m-t_m}\right)q_f +\theta_m =\frac{\phi_{mf}-\psi_{mf}}{L_f}, \quad \forall f\in\mathcal F.
\end{equation}
Due to the fact that $\frac{ b_m}{w_m} +\frac{D}{C_m-t_m}>0$ and $q_1>q_2>\cdots>q_F>0$, we can obtain
\begin{equation}\label{appenBeq3_2}
\frac{\phi_{m1}-\psi_{m1}}{L_1}<\frac{\phi_{m2}-\psi_{m2}}{L_2}<
\cdots<\frac{\phi_{mF}-\psi_{mF}}{L_F}.
\end{equation}
Since $L_f>0$ for all $f\in\mathcal F$, we consider the following four cases.
\begin{enumerate}
  \item If $\frac{\phi_{m1}-\psi_{m1}}{L_1}>0$, we have $\phi_{mf}-\psi_{mf}>0$ for all $f\in\mathcal F$ according to (\ref{appenBeq3_2}).
      Since $\psi_{mf}\geq0$, we further can obtain $\phi_{mf}>0$ for all $f\in\mathcal F$.
      Based on the complementary slackness condition $\phi_{mf}(-s_{mf})=0$, we have $s_{mf}=0$ for all $f\in\mathcal F$.
  \item If $\frac{\phi_{mF}-\psi_{mF}}{L_F}<0$, we have $\phi_{mf}-\psi_{mf}<0$ for all $f\in\mathcal F$ according to (\ref{appenBeq3_2}).
      Since $\phi_{mf}\geq0$, we further can obtain $\psi_{mf}>\phi_{mf}\geq0$ for all $f\in\mathcal F$.
      Based on the complementary slackness condition $\psi_{mf}(1-s_{mf})=0$, we have $s_{mf}=1$ for all $f\in\mathcal F$.
  \item If there exists one $f\in\mathcal F$ such that $\phi_{mf}-\psi_{mf}=0$.
      For $l\in\{1, \cdots, f-1\}$, we have $\frac{\phi_{ml}-\psi_{ml}}{L_l}<\frac{\phi_{mf}-\psi_{mf}}{L_f}=0$ from (\ref{appenBeq3_2}).
      Considering $\phi_{ml}\geq 0$ and the complementary slackness condition, we can obtain $s_{mf}=1$ for all $l\in\{1, \cdots, f-1\}$.
      For $l\in\{f+1, \cdots, F\}$, we have $\frac{\phi_{ml}-\psi_{ml}}{L_l}>\frac{\phi_{mf}-\psi_{mf}}{L_f}=0$ from (\ref{appenBeq3_2}).
      Considering $\psi_{ml}\geq 0$ and the complementary slackness condition, we can obtain $s_{mf}=0$ for all $l\in\{f+1, \cdots, F\}$.
      \item If there exists one $f\in\mathcal F$ such that $\phi_{mf}-\psi_{mf}<0$ and $\phi_{m(f+1)}-\psi_{m(f+1)}>0$.
      For $l\in\{1, \cdots, f\}$, we have $\frac{\phi_{ml}-\psi_{ml}}{L_l}<\frac{\phi_{mf}-\psi_{mf}}{L_f}<0$ from (\ref{appenBeq3_2}).
      Considering $\phi_{ml}\geq 0$ and the complementary slackness condition, we can obtain $s_{mf}=1$ for all $l\in\{1, \cdots, f\}$.
      For $l\in\{f+1, \cdots, F\}$, we have $\frac{\phi_{ml}-\psi_{ml}}{L_l}\geq\frac{\phi_{m(f+1)}-\psi_{m(f+1)}}
      {L_{f+1}}>0$ from (\ref{appenBeq3_2}).
      Considering $\psi_{ml}\geq 0$ and the complementary slackness condition, we can obtain $s_{mf}=0$ for all $l\in\{f+1, \cdots, F\}$.
\end{enumerate}
Based on the above analysis, the optimal solution of Problem (\ref{appenBeq1}) with any given $t_m$ has the structure ($\pmb 1_{f-1}, s_{mf}^*, \pmb 0_{F-f}$), where
\begin{equation}\label{appenBeq5}
\pmb 1_{f-1}=[\underbrace{1, \cdots, 1}_{f-1}], \pmb 0_{F-f}=[\underbrace{0, \cdots, 0}_{F-f}],
\end{equation}
 $s_{mf}^*\in[0,1]$, and $f\in\mathcal F$.
Since Problem (\ref{OCPmin2}) is equivalent to Problem (\ref{appenBeq1}), Theorem 1 is proved.

\section{Proof of Theorem 2}
\setcounter{equation}{0}
\renewcommand{\theequation}{\thesection.\arabic{equation}}

Based on Theorem 1, the optimal solution of Problem (\ref{OCPmin2}) has the structure ($\pmb 1_{f-1}, s_{mf}^*, \pmb 0_{F-f}$) with $s_{mf}^*\in[0,1]$ and $f\in\mathcal F$.
As a result, the optimal solution of Problem (\ref{OCPmin2}) is one the $F$ solutions, ($s_{m1}^*, \pmb 0_{F-1}$), ($1, s_{m2}^*, \pmb 0_{F-2}$), $\cdots$, ($\pmb 1_{F-1}, s_{mF}^*$), with the best objective value.
For the $f$-th solution ($\pmb 1_{f-1}, s_{mf}^*, \pmb 0_{F-f}$),  constraint (\ref{OCPmin2}b) should be satisfied, i.e., $\sum_{l=1}^{f-1}L_l \leq C_m$ and then $s_{mf}^*$ can be obtained by substituting the optimal values of other $F-1$ variables into Problem (\ref{OCPmin2}), i.e., $s_{mf}^*$ is the optimal solution of the following problem:
\begin{subequations}\label{appenCeq1}
\begin{align}
\mathop{\min}_{{s}_{mf}}& \:\;
\left(\frac{ b_m}{w_m} +\frac{D}
{C_m-\sum_{l=1}^{f-1}L_l-s_{mf}L_f}\right)\left(q_f L_f(1-s_{mf})+\sum_{l=f+1}^Fq_lL_l\right)
\triangleq g_{mf}(s_{mf})
\\
\textrm{s.t.}
& \:\; 0\leq s_{mf}\leq s_{mf}^{\max},
\end{align}
\end{subequations}
where
\begin{equation}\label{apenCeq0}
s_{mf}^{\max}=\min\left\{1,\frac{C_m-\sum_{l=1}^{f-1}L_l}{L_f}\right\}.
\end{equation}
The first-order derivative of the objective function (\ref{appenCeq1}a) is
\begin{equation}\label{appenCeq2}
g_{mf}'(s_{mf})=\frac{DL_f\left(q_f L_f(1-s_{mf})+\sum_{l=f+1}^Fq_lL_l\right)}
{\left(C_m-\sum_{l=1}^{f-1}L_l-s_{mf}L_f\right)^2}
-q_fL_f\left(\frac{ b_m}{w_m} +\frac{D}
{C_m-\sum_{l=1}^{f-1}L_l-s_{mf}L_f}\right).
\end{equation}
Setting the first-order derivative (\ref{appenCeq2}) with 0 yields
\begin{equation}\label{appenCeq2_2}
-b_m q_f\left(C_m-\sum_{l=1}^{f-1}L_l-s_{mf}L_f\right)^2
+w_mD\left(-q_fC_m+\sum_{l=f+1}^Fq_lL_l+\sum_{l=1}^{f}q_fL_l\right)=0.
\end{equation}
To solve (\ref{appenCeq2_2}), we consider the following two cases.
\begin{enumerate}
  \item If $-q_fC_m+\sum_{l=f+1}^Fq_lL_l+\sum_{l=1}^{f}q_fL_l\leq0$, the left term of equation (\ref{appenCeq2_2}) is always nonpositive, i.e., $g_{mf}'(s_{mf})\leq 0$ for all $s_{mf}\geq 0$.
       The objective function $g_{mf}(s_{mf})$ monotonically decreases with $s_{mf}$, and the optimal $s_{mf}^*=s_{mf}^{\max}$.
  \item If $-q_fC_m+\sum_{l=f+1}^Fq_lL_l+\sum_{l=1}^{f}q_fL_l>0$, there exists two different roots to equation (\ref{appenCeq2_2}), i.e.,
      \begin{equation}\label{appenCeq2_5}
      s_{mf}(1)=\frac{(C_m-\sum_{l=1}^{f-1}L_l)\sqrt{b_m q_f}-
      \sqrt{w_mD\left(-q_fC_m+\sum_{l=f+1}^Fq_lL_l+\sum_{l=1}^{f}q_fL_l\right)}}
      {L_f\sqrt{b_m q_f}},
      \end{equation}
      and
      \begin{equation}\label{appenCeq2_6}
      s_{mf}(2)=\frac{(C_m-\sum_{l=1}^{f-1}L_l)\sqrt{b_m q_f}+
      \sqrt{w_mD\left(-q_fC_m+\sum_{l=f+1}^Fq_lL_l+\sum_{l=1}^{f}q_fL_l\right)}}
      {L_f\sqrt{b_m q_f}}.
      \end{equation}
      Since the objective function $g_{mf}(s_{mf})$ decreases with $s_{mf}$ when $s_{mf}<s_{mf}(1)$ and $s_{mf}>s_{mf}(2)$ and increases with $s_{mf}$ when $s_{mf}(1)\leq s_{mf}\leq s_{mf}(2)$.
      If $0<s_{mf}(1)<s_{mf}^{\max}$, we have the optimal $s_{mf}^*=\arg\min_{s_{mf}\in\{s_{mf}(1),s_{mf}^{\max}\}}g_{mf}(s_{mf})$.
      If $s_{mf}(1)\leq 0$, $s_{mf}^*=\arg\min_{s_{mf}\in\{0,s_{mf}^{\max}\}}g_{mf}(s_{mf})$.
      If $s_{mf}(1) \geq s_{mf}^{\max}$, we have the optimal $s_{mf}^*=s_{mf}^{\max}$.
\end{enumerate}

\section{Proof of Theorem 3}
\setcounter{equation}{0}
\renewcommand{\theequation}{\thesection.\arabic{equation}}

Denoting by $\chi$ the Lagrange multiplier associated to (\ref{OBAmin1}b), the Lagrange function of Problem (\ref{OBAmin1}) is
\begin{equation}\label{apenAeq1}
\mathcal L_2(\pmb w, \chi)
=
\frac{ \sum_{f\in\mathcal F}q_f L_f \sum_{m\in\mathcal M\cup\{0\}}a_m}{w_0}+
\sum_{m\in\mathcal M}
\frac{\sum_{f\in\mathcal F}b_m q_f L_f (1-s_{mf})}{w_m}
+\chi\left( \sum_{m\in\mathcal M\cup\{0\}} w_m - W\right).
\end{equation}
The optimal solution should satisfy the following KKT conditions of Problem (\ref{OBAmin1}):
\begin{subequations}\label{appenAKKT1}
\begin{align}
&
\frac {\partial \mathcal L_2}{\partial w_0}=
-\frac{ \sum_{f\in\mathcal F}q_f L_f \sum_{m\in\mathcal M\cup\{0\}}a_m}{w_0^2}
+\chi=0
\\
&\frac {\partial \mathcal L_2}{\partial w_m}=
-\frac{\sum_{f\in\mathcal F}b_m q_f L_f (1-s_{mf})}{w_m^2}
+\chi=0,\quad\forall m\in\mathcal M,
\end{align}
\end{subequations}
which yields
\begin{equation}\label{apenAeq2_1}
w_0=\frac{ \sqrt{\sum_{f\in\mathcal F}q_f L_f \sum_{m\in\mathcal M\cup\{0\}}a_m}}{\sqrt\chi},
\end{equation}
and
\begin{equation}\label{apenAeq2_2}
w_m=\frac{\sqrt{\sum_{f\in\mathcal F}b_m q_f L_f (1-s_{mf})}}{\sqrt\chi}, \quad\forall m\in \mathcal M.
\end{equation}
Substituting (\ref{apenAeq2_1}) and (\ref{apenAeq2_2}) into (\ref{OBAmin1}b), we can obtain
\begin{equation}\label{apenAeq3}
\sqrt\chi=\frac{\sqrt{ \sum_{f\in\mathcal F}q_f L_f \sum_{m\in\mathcal M\cup\{0\}}a_m}+\sum_{m\in\mathcal M}\sqrt{\sum_{f\in\mathcal F}b_m q_f L_f (1-s_{mf})}}{W}.
\end{equation}
Substituting (\ref{apenAeq3}) into (\ref{apenAeq2_1}) and (\ref{apenAeq2_2}) yields (\ref{OBAeq1}) and (\ref{OBAeq1_2}), respectively.

\section{Proof of Theorem 4}
\setcounter{equation}{0}
\renewcommand{\theequation}{\thesection.\arabic{equation}}

Introducing auxiliary variable $\pmb t=[t_1, \cdots, t_M]^T$, Problem (\ref{JBCmin1}) is equivalent to
\begin{subequations}\label{appenDeq1}
\begin{align}
\mathop{\min}_{ \pmb{s}, \pmb t}& \:\;
\frac1W \left( a +\sum_{m\in\mathcal M}\sqrt{\sum_{f\in\mathcal F}b_m q_f L_f (1-s_{mf})}\right)^2
+\sum_{m\in\mathcal M}
\frac{ \sum_{f\in\mathcal F}q_fL_fD(1-s_{mf})}
{C_m-t_m}
\\
\textrm{s.t.}
& \:\;\sum_{f\in\mathcal F}L_f s_{mf}\leq t_m, \quad \forall m \in \mathcal M\\
& \:\;t_m\leq C_m, \quad \forall m \in \mathcal M\\
& \:\; 0\leq s_{mf}\leq 1,\quad\forall m\in\mathcal M, f \in \mathcal F.
\end{align}
\end{subequations}
With given $\pmb t$, the Lagrangian function of Problem (\ref{appenDeq1}) is
\begin{eqnarray}\label{apenBeq2}
\mathcal L_3\!\!\!\!\!\!&&\!\!\!(\pmb s, \pmb \theta, \pmb \phi, \pmb \psi)
=\frac1W \left( a +\sum_{m\in\mathcal M}\sqrt{\sum_{f\in\mathcal F}b_m q_f L_f (1-s_{mf})}\right)^2
+\sum_{m\in\mathcal M}
\frac{ \sum_{f\in\mathcal F}q_fL_fD(1-s_{mf})}
{C_m-t_m}
\nonumber\\
&&\:\;
+ \sum_{m\in\mathcal M} \theta_m\left(\sum_{f\in\mathcal F}L_fs_{mf}-t_m\right)
+\sum_{m\in\mathcal M}\sum_{f\in\mathcal F}\phi_{mf}(-s_{mf})+\sum_{m\in\mathcal M}\sum_{f\in\mathcal F}\psi_{mf}(s_{mf}-1),
\end{eqnarray}
where $\pmb\theta=[\theta_1, \cdots, \theta_M]^T$, $\pmb \phi=[\phi_{11},\cdots,\phi_{1F},\cdots,\phi_{MF}]^T$ and $\pmb \psi=[\psi_{11},\cdots,\psi_{1F},\cdots,\psi_{MF}]^T$ are nonnegative Lagrangian multipliers associated with corresponding  constraints of Problem (\ref{appenDeq1}).
The optimal solution should satisfy the following KKT conditions of Problem (\ref{appenDeq1}):
\begin{equation}\label{appenDKKT1}
\frac {\partial \mathcal L_3}{\partial s_{mf}}=
-\left(\frac{ ab_m+\sum_{n\in\mathcal M} b_mv_n}{Wv_m} +\frac{D}{C_m-t_m}\right)q_fL_f+\theta_m L_f-\phi_{mf}+\psi_{mf}=0, \quad \forall m\in\mathcal M, f\in\mathcal F,
\end{equation}
where $v_m=\sqrt{\sum_{f\in\mathcal F}q_f L_f b_m(1-s_{mf})}$, $\forall m \in \mathcal M$.
From (\ref{appenBKKT1}), we have
\begin{equation}\label{appenDeq3}
-\left(\frac{ ab_m+\sum_{n\in\mathcal M} b_mv_n}{Wv_m} +\frac{D}{C_m-t_m}\right)q_f +\theta_m =\frac{\phi_{mf}-\psi_{mf}}{L_f}, \quad \forall m\in\mathcal M, f\in\mathcal F.
\end{equation}
Due to the fact that $\frac{ ab_m+\sum_{n\in\mathcal M} b_mv_n}{Wv_m} +\frac{D}{C_m-t_m}>0$ and $q_1>q_2>\cdots>q_F>0$, we can obtain
\begin{equation}\label{appenDeq3_2}
\frac{\phi_{m1}-\psi_{m1}}{L_1}<\frac{\phi_{m2}-\psi_{m2}}{L_2}<
\cdots<\frac{\phi_{mF}-\psi_{mF}}{L_F}, \quad \forall m\in \mathcal M.
\end{equation}
Similar to the analysis in Appendix B, the optimal solution of Problem (\ref{appenDeq1}) with any given $\pmb t$ must have the structure
\begin{equation}
(\pmb 1_{f_1-1}, s_{1f_1}^*, \pmb 0_{F-f_1},\cdots,\pmb 1_{f_m-1}, s_{mf_m}^*, \pmb 0_{F-f_m},\cdots,\pmb 1_{f_M-1}, s_{Mf_M}^*, \pmb 0_{F-f_M}),
\end{equation}
where $s_{mf_m}^*\in[0,1]$, $f_m\in\mathcal F$, $m\in\mathcal M$.
Since Problem (\ref{OCPmin2}) is equivalent to Problem (\ref{appenDeq1}), Theorem 4 is proved.

\section{Proof of Theorem 5}
\setcounter{equation}{0}
\renewcommand{\theequation}{\thesection.\arabic{equation}}

According to Theorem 4, the optimal $\pmb s_m$ for pico BS $m$ to Problem (\ref{JBCmin1}) with given $\pmb s_{-m}$ is one of the $F$ solutions, ($s_{m1}^*, \pmb 0_{F-1}$), ($1, s_{m2}^*, \pmb 0_{F-2}$), $\cdots$, ($\pmb 1_{F-1}, s_{mF}^*$), with the best objective value.
For the $f$-th solution ($\pmb 1_{f-1}, s_{mf}^*, \pmb 0_{F-f}$),  $\sum_{l=1}^{f-1}L_l \leq C_m$ should be first satisfied from (\ref{JBCmin1}b) and $s_{mf}^*$ is the optimal solution of the following problem according to (\ref{JBCmin1}):
\begin{subequations}\label{appenEeq1}
\begin{align}
\!\!\mathop{\min}_{{s}_{mf}}& \:\;
\frac1W (u_m + \sqrt{b_m w_{f} -b_mq_fL_fs_{mf}})^2
+
\frac{w_{f}D-q_fL_fDs_{mf}}
{C_m-\sum_{l=1}^{f-1}L_l-L_f s_{mf}}
\triangleq y_{mf}(s_{mf})
\\
\textrm{s.t.}
& \:\; 0\leq s_{mf}\leq s_{mf}^{\max},
\end{align}
\end{subequations}
where $u_m=a +\sum_{n\in\mathcal M\setminus \{m\}}\sqrt{\sum_{l\in\mathcal F}b_n q_l L_l (1-s_{nl})}$,
$w_{f}=\sum_{l=f}^Fq_l L_l$,
$s_{mf}^{\max}$ is defined in (\ref{apenCeq0}).
The first-order derivative of the objective function (\ref{appenEeq1}a) is
\begin{equation}\label{appenEeq2}
y_{mf}'(s_{mf})=
\frac{-b_m q_fL_f(u_m + \sqrt{b_m w_{f}-b_mq_fL_fs_{mf}})} {W\sqrt{b_m w_{f}-b_mq_fL_fs_{mf}}}
+\frac{\left(w_{f}-q_f C_m+q_f\sum_{l=1}^{f-1}L_l\right)L_fD}
{\left(C_m-\sum_{l=1}^{f-1}L_l-L_f s_{mf}\right)^2}.
\end{equation}
Setting $y_{mf}'(s_{mf})=0$ and $x=\sqrt{b_m w_{f}-b_mq_fL_fs_{mf}}$ to (\ref{appenEeq2}) yields
\begin{equation}\label{appenEeq2_2}
z_5 x^5+z_4x^4+z_3x^3+z_2x^2+z_1x+z_0=0,
\end{equation}
where $z_5=b_m q_fl_f$, $z_4=b_mu_m q_fl_f$, $z_3=-2b_m^2 q_fl_f \left(q_fC_m-q_f\sum_{l=1}^{f-1}L_l-w_f\right)$,
$z_2=2b_m^2 u_m q_fl_f$ $\left(q_fC_m-q_f\sum_{l=1}^{f-1}L_l-w_f\right)$,
$z_1=b_m^3 q_fl_f\left(q_fC_m-q_f\sum_{l=1}^{f-1}L_l-w_f\right)^2
-b_m^2q_f^2L_fDW\Big(w_{f}-$ $q_f C_m+q_f\sum_{l=1}^{f-1}L_l\Big)$,
$z_0=b_m^2u_m\left(q_fC_m-q_f\sum_{l=1}^{f-1}L_l-w_f\right)^2$.

Having obtained $x$ from (\ref{appenEeq2_2}), $s_{mf}$ can be presented by $s_{mf}=\frac{b_m w_f -x^2}{b_m q_f L_f}=\frac{b_m \sum_{l=f} ^F q_lL_l-x^2}{b_m q_f L_f}$.
Due to that $s_{mf}\in(0,s_{mf}^{\max})$, $x$ should be in the interval $(\sqrt{b_m \sum_{l=f} ^F q_lL_l-b_mq_fL_fs_{mf}^{\max}}, \sqrt{b_m \sum_{l=f} ^F q_lL_l})$.
According to Abel-Ruffini theorem \cite{pesic2003abel}, there is no algebraic expression for general quintic equations over the rationals in terms of radicals.
The roots located in $(\sqrt{b_m \sum_{l=f} ^F q_lL_l-b_mq_fL_fs_{mf}^{\max}},$ $\sqrt{b_m \sum_{l=f} ^F q_lL_l})$ to equation are numerically calculated using root-finding algorithm for polynomials \cite{madsen1973root}.
Since the optimal solution of Problem (\ref{appenEeq1}) either lies in the boundary or in the extreme point, the optimal $s_{mf}^*$ can be presented in (\ref{JBCeq1}).

\bibliographystyle{IEEEtran}
\bibliography{IEEEabrv,MMM}

\end{document}